\documentclass{aa}
\usepackage{graphicx}
\def \ha   {H$\alpha$}
\def \oiii {[O~{\sc iii}]5007~\AA}

\def \NII  {[N~{\sc ii}]~6548~\&~6584~\AA}
\def \sii {[S~{\sc ii}]~6717 \& 6731\AA}

\def \tena#1 #2 {\ifmmode{#1 \times 10^{#2}}\else{$#1 \times 10^{#2}$}\fi}
\def \kms  {\ifmmode{~{\rm km\,s}^{-1}}\else{~km s$^{-1}$}\fi}
\def \vhel {\ifmmode{~V_{{\rm HEL}}}\else{~$V_{{\rm HEL}}$}\fi}
\def \vlsr {\ifmmode{~V_{{\rm LSR}}}\else{~$V_{{\rm LSR}}$}\fi}
\def \vsys {\ifmmode{~V_{{\rm sys}}}\else{~$V_{{\rm sys}}$}\fi}
\def \vexp {\ifmmode{~V_{{\rm exp}}}\else{~$V_{{\rm exp}}$}\fi}

\def \deg  {\ifmmode{^{\circ}}\else{$^{\circ}$}\fi} 
\def \msun {\ifmmode{{\rm\ M}_\odot}\else{${\rm\ M}_\odot$}\fi}
\def \myr  {\ifmmode{{\rm\ M}_\odot{\rm\ yr}^{-1}}
         \else{${\rm\ M}_\odot$ yr$^{-1}$}\fi}
\def\mnras{MNRAS}
\def\apj{ApJ}
\def\apjs{ApJS}
\def\aap{A\&A}

\def\aj{AJ}
\def\pasp{PASP}

\begin{document}
\title{A high-speed bi-polar outflow from the archetypical pulsating
star Mira A}
\author{J. Meaburn\inst{1}
   \and J. A. L{\'o}pez\inst{2}
   \and P. Boumis\inst{3}
   \and M. Lloyd\inst{1}
   \and M. P. Redman\inst{4}}
\offprints{Prof. J. Meaburn}
\authorrunning{J. Meaburn et al.}
\titlerunning{High speed outflows from Mira}
\institute{Jodrell Bank Centre for Astrophysics, University of
Manchester, Manchester, UK, M13 9PL.
\and Instituto de Astronomia, UNAM, Apdo. Postal 877. Ensenada,
B.C. 22800, M\'{e}xico.
\and Institute of Astronomy \& Astrophysics, National Observatory of
Athens, I. Metaxa \& V. Pavlou, P. Penteli, GR-15236 Athens, Greece.
\and Centre for Astronomy, National University of Ireland, Galway,
University Road, Galway, Ireland.}
\date{Received ; accepted }
\abstract{ Optical images and high--dispersion spectra have been
obtained of the ejected material surrounding the pulsating AGB star
Mira A.  The two streams of knots on either side of the star, found in
far ultra--viollet (FUV) GALEX images, have now been imaged clearly in
the light of \ha. Spatially resolved profiles of the same line reveal
that the bulk of these knots form a bi--polar outflow with radial
velocity extremes of $\pm$ 150 \kms\ with respect to the central star.
The South stream is approaching and the North stream receding from the
observer. A displacement away from Mira A between the position of
one of the South stream knots in the new \ha\ image and its position
in the previous Palomar Observatory Sky Survey (POSS I) red plate has
been noted. If interpreted as a consequence of expansion proper
motions the bipolar outflow is tilted at 69\deg\ $\pm$ 2\deg\ to the
plane of the sky, has an outflow velocity of 160 $\pm$ 10 \kms\ and is
$\approx$ 1000 y old.
}
\maketitle
\section{Introduction}
The binary system Mira AB has attracted attention over 400 yr (for a
summary see Hoffleit, 1997), for apart from being very close to the
Sun (the Hipparcos distance is 107 $\pm$ 10 pc) Mira A is the
archetypical pulsating variable star with a period of hundreds of days
and brightness range of 8 mag (Reid \& Goldston 2002). Its periodic
outbursts over the last 80 cycles are listed by Percy \& Bagby (1999).
Hubble Space telescope images have shown Mira B to be separated by
$\approx$ 0.5\arcsec\ from A ($\equiv$ 54 AU; Matthews \& Karovska
2006). Mira A of intermediate mass is certainly in its volatile
asymptotic giant branch (AGB) phase whereas B was previously
identified as a White Dwarf but recently as a somewhat innocuous
F-type star similar to the Sun though with an accretion disk formed by
the mass ejections from A (Karkova et al. 2005; Ireland et al. 2007
and refs. cited therein).  In any case B is too well separated from A
to affect significantly the overall dynamics of the ejected material.

Interest in Mira A has been further heightened by the discovery with
the GALEX satellite of a 2\deg\ long ($\approx$ 4 pc) FUV emitting
tail (Martin et al. 2007). They propose that this traces the
interactions with the local interstellar medium (ISM) over the past
30,000~y of the ejected material from Mira A. This is plausible
(Wareing et al. 2007) for the star has a high space velocity of 120
\kms\ at an angle of 28\deg\ with respect to the plane of the sky
tilted away from the observer and what appears to be a preceding
bow-shock (Raga \& Canto 2008).  Perhaps relevant is the tail of
similar dimensions apparently projecting from the similarly
mass--ejecting proto--typical Luminous Blue Variable (LBV) star
P~Cygni (Meaburn et al.  1999); though later, deeper, observations in
the higher excitation \oiii\ line show this tail as part of a larger
complex of filaments but still convincingly associated with the star
(Meaburn et al. 2004; Boumis et al. 2006 and references therein).

In the GALEX FUV images a group of emission knots close to Mira AB and
called North Stream and South Stream by Martin et al. (2007) have been
identified as recent manifestations of the mass ejections from Mira
A. As these are within a field of a few arcmin diameter they have now
been investigated at optical wavelengths using occulting strip
imagery, to exclude scattered continuum light from the bright stellar
image, and longslit, spatially resolved, high-dispersion
spectrometry. The results of these optical observations will now be
presented.
\section{Observations and results}
Observations were made with the Manchester Echelle Spectrometer
(MES-SPM - see Meaburn et al. 1984; 2003) combined with the 2.1-m San
Pedro Martir, (B.C. Mexico) telescope on 2008, 26 to 28 Oct.  A
SITe CCD was the detector with 1024$\times$1024, 24~$\mu$m pixels
although 2$\times$2 binning was employed.  The `seeing' throughout was
$\approx$~1\arcsec.
\subsection{Imagery}
MES-SPM has a limited imaging capability with a 
retractable plane mirror isolating
the echelle grating and a clear area (4.37$\times$5.32
arcmin$^{2}$) replacing the spectrometer slit.

The image in Fig. 1 is a subset from this larger field obtained on the
2008 28 Oct. An occulting strip of chromium suppressed the bright
image of Mira.  The integration time was 2000~s.  The coordinates
(J2000) were added using the STARLINK \textsc{gaia} software. The
90\AA\ bandwidth interference filter transmits predominantly the \ha\
nebular line as \NII\ emission has been shown (Martin et al. 2007) to
be very low.  Confusing, faint star images were eliminated using the
PATCH routine of the STARLINK \textsc{gaia} software.  Deep and
lighter representations of the new \ha\ image are compared in Fig. 1
to the FUV GALEX image from Martin et al. (2007).
\subsection{Longslit spectroscopy} 
Spatially resolved, longslit \ha\ line profiles were obtained with the
MES--SPM along the lines marked 1--6 in Fig. 2. These are only partial
lengths of the full NS slits as measurable \ha\ emission only occurred
over small sections of the full slit lengths containing the bright
knots in the North and South streams. The increments along the slit
length each corresponds to 0.63\arcsec.
 
In this spectroscopic mode MES--SPM has no cross--dispersion
consequently, for the present observations, a filter of 90~\AA\
bandwidth was used to isolate the 87$^{\rm th}$ echelle order
containing the \ha\ and \NII\ nebular emission lines. The slit width
was always 300 $\mu$m which is $\equiv$ 3.8\arcsec\ on the sky and 20
\kms\ spectral halfwidth (HPBW).  Each integration time was 1800 s.
Also indicated in Fig. 2 are the directions of the 2\deg\ long FUV
emitting tail detected by Martin et al. (2007) and of the proposed
bowshock found in the same wavelength domain.

The longslit spectra were cleaned of cosmic rays, bias corrected and
calibrated in wavelength to $\pm$~1~\kms\ accuracy in the usual way
using STARLINK \textsc{figaro} software. The greyscale representation
of the position--velocity (pv) arrays of \ha\ line profiles (after
subtraction of the sky background spectra) for the partial slit
lengths shown in Fig. 2 for Slits 1--4 and 5--6 are shown in Figs. 3
\& 4 respectively. Here contours of the \ha\ relative surface
brightnesses with linear intervals have been overlain on the negative
greyscale representations.  As no standard star was observed the
absolute surface brightnesses are unreliable and will not be presented
here. Line profiles from various incremental lengths of these pv
arrays of profiles are shown in Fig. 5.
\section{Discussion}
\subsection{The bi-polar outflow}
The radial velocities of the South stream of knots shown in Fig. 3
show a shift of around -150 \kms\ with respect to the heliocentric
systemic radial velocity \vsys\ = 56 \kms\ as given by Matthews et
al. (2008 -- N.B. \vlsr\ = \vhel\ - 9.44 \kms\ for Mira A).  A very
similar shift, but now to positive radial velocities can be seen in
Fig. 4 for the North Stream on the opposite side of Mira A. Only the
more distant arc from Mira A (Figs. 1 \& 2) covered by the element
Slit 5a (see the profile in Fig. 5) has a greater shift (200 \kms) in
radial velocity with respect to \vsys. There seems therefore to be a
strong bi-polar structure to the principal elements of this outflow.

This expansion radial velocity measurement of the bi-polar lobes
has to be combined with a measurement of their expansion tangential
velocity relative to Mira A to determine their actual expansion
velocity. The expansion proper motions (PMs) then of the knots
relative to Mira A are required for this purpose. Simple inspection of
the \ha\ image in Fig. 1b and FUV image in Fig. 1c reveal that the
knots in \ha\ emission are $\approx$ 2\arcsec\ closer to the reference
star at the bottom of both frames than in the FUV emission. This could
indicate the displacement caused by a high expansion PM in the
intervening 1.9 y between the two observation leading to an expansion
velocity of $\approx$~600\kms. However, a more likely cause is the
spatial difference within the knots of the \ha\ and FUV emission
regions.

This suggestion is born out by the spatial displacement of Knot B
(Fig. 2) in the first Palomar Observatory Sky Survey POSS I red plate
taken on the 26 Nov. 1954 and the \ha\ image in Fig. 1b obtained on
the 20 Oct 2008. Only Knot B is clearly identifiable on this POSSI
plate with the image of the brighter Knot A confused with that of a
faint star.Curiously none of the Mira knots nor this star are
detectable on the follow-up POSS II (23 Aug 1995) red plate.  However,
using the STARLINK \textsc{gaia} software the displacement of Knot B,
measured with respect to the faint star field, in the intervening
53.90 y is +2\arcsec\ in Right Ascension ($\alpha$) and -17\arcsec in
Declination ($\delta$) with an uncertainty of $\pm$2\arcsec\ in both
measurements determined mainly by the poor quality of the image of
Knot B in the POSS I plate.  This low quality is not surprising for
the POSS bandwidth is many times greater than the present one,
photographic emulsion was the detector and the focal length
shorter. The PM of Knot B is then measured as d($\alpha$,$\delta$)/dt
= (+37, -319) mas y$^{-1}$ each to around $\pm$40 mas y$^{-1}$
uncertainty. Here it is assumed that \ha\ dominates in the POSS I
detection even though the bandwidth also encompasses the \sii\ lines
and much more continuum light than the present observations.

For comparison a similar PM measurement of Mira A itself was made
between the POSS I and POSS II red plates. In the intervening 40.74 y
the displacement is -1.7\arcsec\ in right ascension and -8.4\arcsec\
in declination to give d($\alpha$,$\delta$)/dt = (-40, -206) mas
y$^{-1}$ with ($\pm$30) mas y$^{-1}$ uncertainties for the Mira A
positions for these were only determined from the centroids of the
diffraction spikes in the two images but still corrected for the small
differences in coordinates of the faint star field.

It is firstly interesting that from these PM measurements Knot B is
moving along PA = 173\deg\ which is reasonably aligned with bipolar
axis of the South stream knots in Fig. 1 and tilted by a few degrees
to the direction of Mira A (as noted by Martin et al. 2007) along PA
$\approx$ 190\deg\ which is also the orientation of the long FUV tail
(Fig. 2).  An expansion along this bi-polar axis of $\approx$ 115 mas
y$^{-1}$ of Knot B relative to Mira A is therefore indicated to give
an expansion tangential velocity of 58 $\pm$ 20\kms. When combined
with the expansion radial velocity of 150 \kms, this indicates an
outflow velocity of $\approx$160 \kms\ tilted for the southern lobe at
69 $\pm$ 15\deg\ towards the observer with respect to the plane of the
sky when all uncertainties are considered. More precise PM
measurements of the knots are required to refine these
estimations. The PM of Knot B with respect to Mira A, which is
127\arcsec\ away from it, gives a time of $\approx$ 1000 y since its
ejection. The Mira A outburst in 1070 (Ho 1962; Hoffleit
1997) then become candidate for the creation of these bi--polar
lobes.
\subsection{The leading bow-shock}
The preceding arc of FUV emission in the direction of motion of Mira A
has been interpretated as a bow--shock being generated by the 120
\kms\ space motion of Mira A through the local ISM (Martin et
al. 2007).  It should now be considered that the bi-polar outflow
identified here could contribute to the creation of this
bow--shock. In any case the radial velocity measurements (Figs 3 \& 4)
and the angular extent of the bi-polar lobes (Fig. 1) certainly
indicate, without consideration of expansion PMs, that the outflow
velocity is $\approx$ 200 \kms. Incidently several planetary nebulae
(PNe) with central binary systems, one star of which has evolved
beyond its Mira~A AGB phase, exhibit bow-shocks emitting optical
lines, as they plough through the local ISM (Bond \& Livio 1990 and
refs. therein). However, the question is posed that if bi--polar
ejecta are favoured as the creators of these bow--shocks then why has
GALEX only detected one such system?
\subsection{Evolution}
Mira A is currently in its AGB phase and shows all the signs that it
could evolve soon into a complex PN maybe with multiple bi-polar
lobes. For comparison, such lobes of the PN NGC~ 6302 have been
conclusively shown to have been ejected at a high velocity (630 \kms;
Meaburn et al 2005 \& 2008) and formed over a timescale of $\approx$
2000 yr. The bi--polar outflow from Mira A revealed here could easily
develop into such features over a similar timescale.  Meanwhile, if
Mira~A itself evolves into a hot PN progenitor as it sheds its outer
layers in its current AGB phase during this period these high speed
lobes will become radiatively ionized and emit a wide range of
emission lines.
\begin{acknowledgements}
We acknowledge the excellent support of the staff at the San Pedro
Martir observatory during these observations. JAL gratefully
acknowledges financial support from DGAPA-UNAM grant IN116908. MPR is
supported by the IRCSET, Ireland. We are grateful to the referee,
Dr. B. Welsh, for drawing our attention to the image of southern knots
on the POSS I plates.
\end{acknowledgements}
\begin{figure*}
\centering
\includegraphics[height=0.5\textheight]{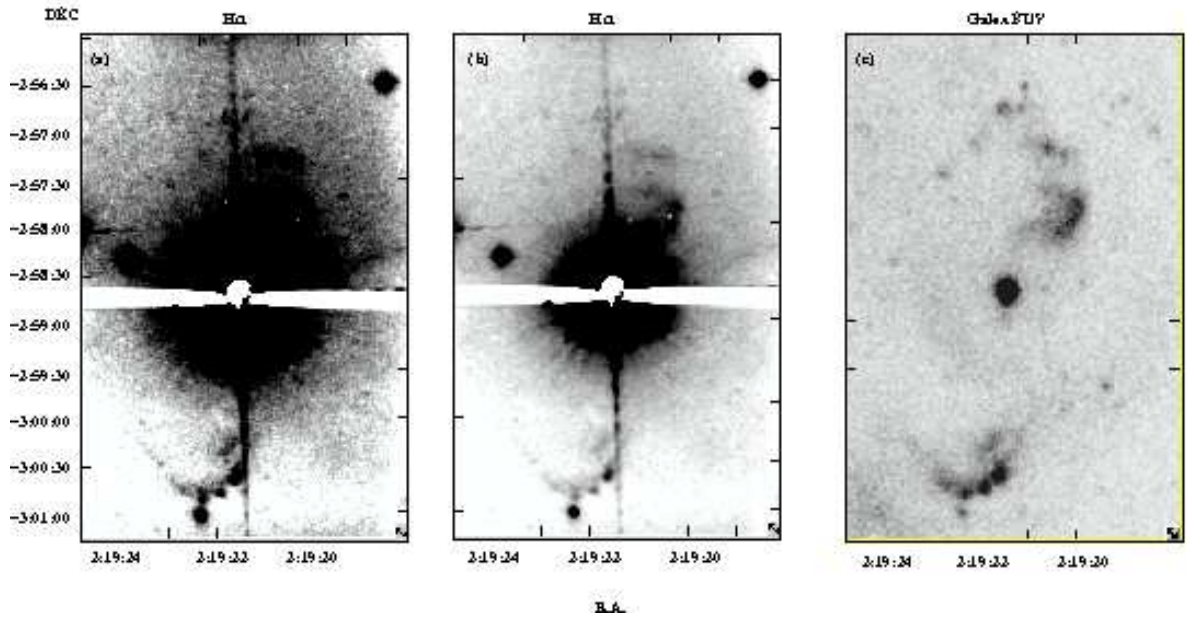}
\caption{Deep and lighter representations in panels a) \& b)
 respectively are of the new \ha\ image of the knots obtained on 2008
 28 Oct. and compared in panel c) with the FUV image of Martin et al.
 (2007) obtained on 2006 18 Nov. -- 15 Dec.  The occulting strip seen
 in a) \& b) minimizes the scattered light of Mira~A within the
 telescope and instrumental optics. Coordinates throughout are 2000
 epoch.  }
\label{Fig1} 
\end{figure*}
\begin{figure*}
\includegraphics[height=0.8\textheight]{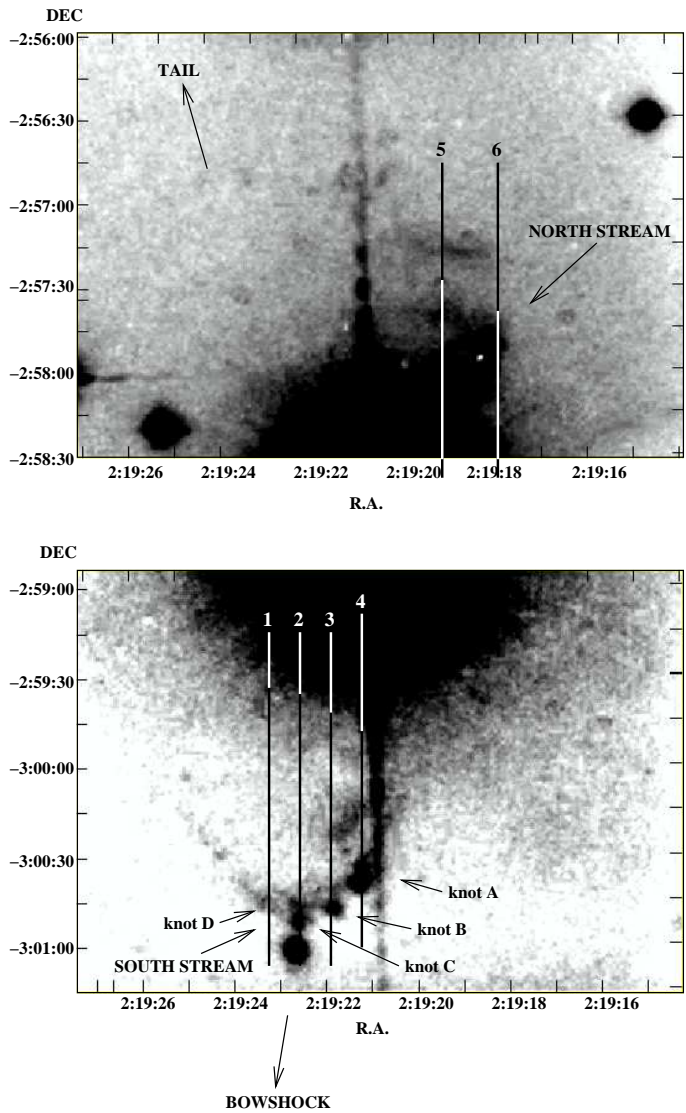}
\caption{Those parts of the lengths of the spectrometer slits
containing useful \ha\ profiles are marked as 1--4 over the southern
knots (South Stream) and 5--6 over the northern knots (North
Sream). The directions of the FUV tail and bow-shock are also
indicated. The top panel is of the new \ha\ image in Fig. 1 but
printed lightly to show the details of the North Stream and the bottom
window is deeper for the South Stream. The relative positions of the
two panels in this figure are somewhat arbitary.}
\label{Fig2} 
\end{figure*}
\begin{figure*}
\includegraphics[height=0.8\textheight]{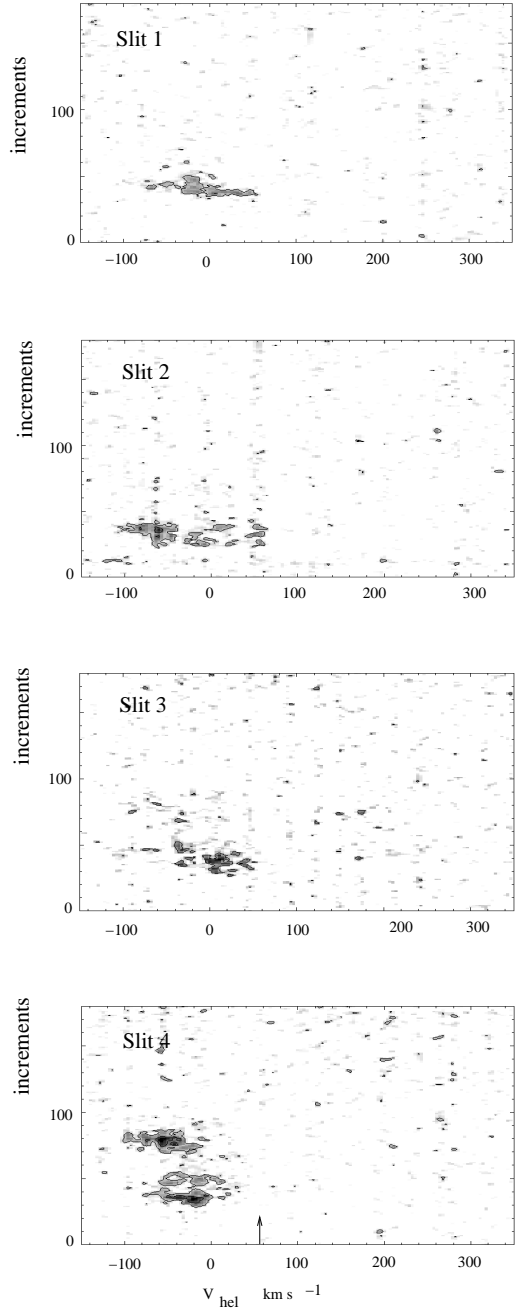}
\caption{Contoured and greyscale representations of the pv arrays of
\ha\ line profiles from the South stream of knots shown in
Figs.1~\&~2. These are for the slit lengths 1 - 4 shown in Fig. 2. The
spectra of the background have been subtracted in all cases. The
heliocentric systemic radial velocity \vsys\ of Mira~A is arrowed.
North is to the top of each slit length and each increment is $\equiv$
0.63 \arcsec.}
\label{Fig3} 
\end{figure*}
\begin{figure*}
\includegraphics[height=0.8\textheight]{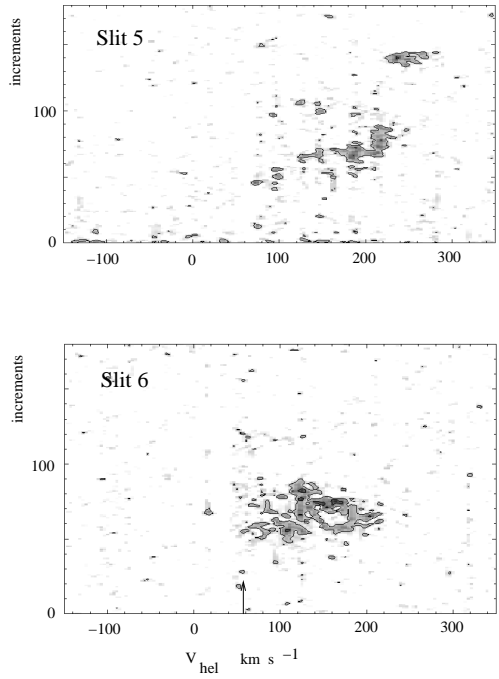}
\caption{As for Fig. 4 but for the northern knots covered by slit
lengths 5 and 6.}
\label{Fig4} 
\end{figure*}
\begin{figure*}
\includegraphics[height=0.8\textheight]{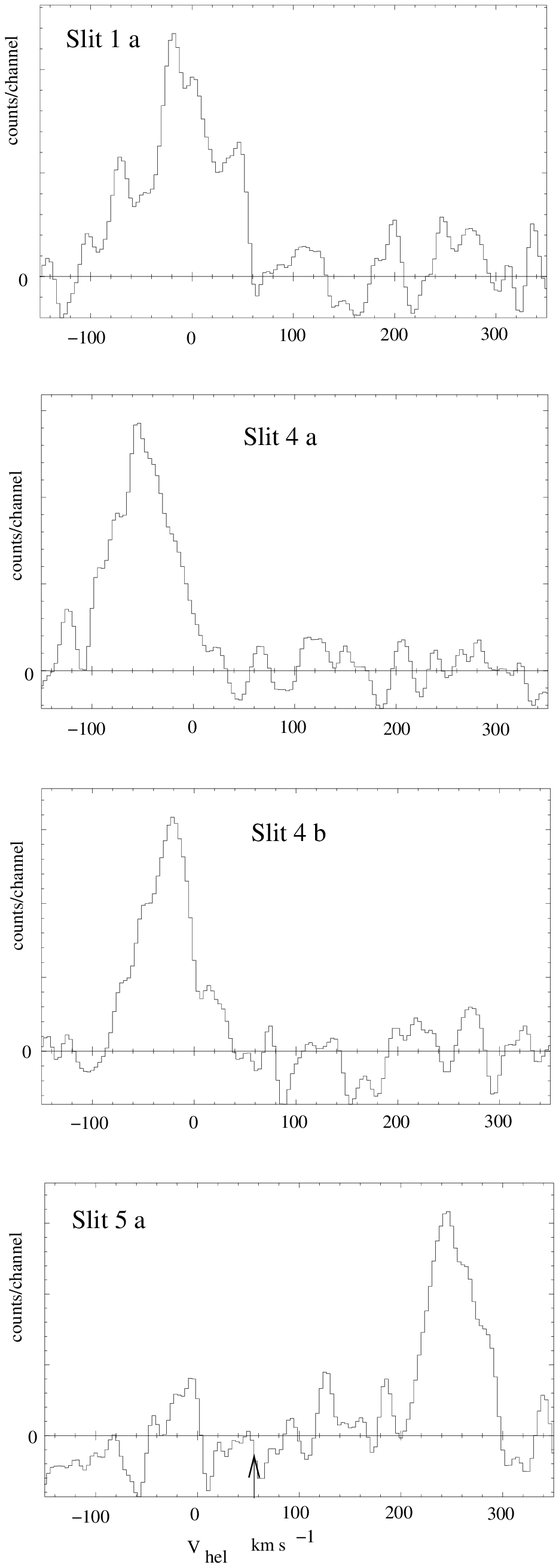}
\caption{Sample line profiles extracted from the pv arrays of line
profiles in Figs. 3 and 4 are shown. That for Slit 1a is for
\ha\ profiles from incremental lengths 30 to 50 coadded along the Slit
1 in Fig. 3 and similarly the profiles Slit 4a and b are for lengths
68 to 83 and 31 to 37 respectively for Slit 4. The profile marked Slit
5a is for increments 132 to 144 along Slit 5 in Fig. 4. As the
background spectra have been subtracted in all of these profiles the
noise fluctuates around the zero level. Again the heliocentric \vsys\
is arrowed. The profiles have been smoothed by a Gaussian of 1.5 times
the 20~\kms\ instrumental HPBW.}
\label{Fig5} 
\end{figure*}

\begin{thebibliography}{}
%
\bibitem[]{}
Bond, H. E. \& Livio, M., 1990, \apj, 355, 568.

\bibitem[]{}
Boumis, P., Meaburn, J., Redman, M. P. \& Mavromatakis, F., 2006, 
\aap, 457, L13.

\bibitem[]{}
Ho, P. Y., 1962, VA, 5, 127

\bibitem[]{}
Hoffleit, D., 1997, J. Am. Ass. Var. Star. Observ., 25, 115.

\bibitem[]{}
Ireland M. J. et al., 2007, \apj, 662, 651.

\bibitem[]{}
Karovska, M., Schiegel, E., Hack, W., Raymond, J. C. \& Wood, B. E. A.,
2005, \apj, 623, L137.

\bibitem[]{}
Martin, D. C., Seibert, M \& Neill, J. D. et al., 2007, Nature, 448, 780.

\bibitem[]{}
Matthews, L. D. \& Karovska, M., 2006, \apj, 637, L49.

\bibitem[]{} 
Matthews, L. D., Libert, Y., G\'{e}rard, E., Le Bertre,
T., Reid, M. J., 2008, ApJ, 684, 603

\bibitem[1984]{Me84}
Meaburn, J., Blundell, B., Carling, R., Gregory, D. E., Keir, D. F. \&
Wynne C. G., 1984, \mnras, 210, 463.

\bibitem[]{} 
Meaburn J., L\'{o}pez J. A. \& O'Connor J. A., 1999, ApJ, 516, L29.

\bibitem[]{}
Meaburn, J., L{\'o}pez, J. A.,  Guti{\'e}rrez, L.,
Quir{\'o}z, F., Murillo, J. M., Vald{\'e}z, J., Pedrayez, M.,  2003,
 RMxAA, 39, 185.

\bibitem[2004]{}
Meaburn,J., Boumis, P., Redman, M. P., L\'{o}pez, J. A \& Mavromatakis, F.,
2004, \aap, 422, 602.

\bibitem[]{}
Meaburn, J., L{\'o}pez, J. A., Steffen, W., Graham, M. F. \& Holloway,
A. J., 2005, \aj, 130, 2303.

\bibitem[]{}
Meaburn, J., Lloyd, M., Vayet, N. M. H. \&  L{\'o}pez, J. A., 2008,
\mnras, 385, 269.

\bibitem[]{}
Percy, J. R. \& Bagby, D. H., 1999, \pasp, 111, 203.

\bibitem[]{}
Raga, A. C. \& Cant{\'o}, J., 2008 \apj, 685, L141.

\bibitem[]{}
Reid, M. J. \& Goldston, J. E., 2001, \apj, 568, 931.

\bibitem[]{}
Wareing, C. J., Zijlstra, A. A., O'Brien, T. J. \& Seibert, M., 2007,
\apj, 670, L125.
%
\end{thebibliography}
\end{document}